\documentclass[superscriptaddress,twocolumn,floatfix,showkeys,preprintnumbers,showpacs]{revtex4}
\usepackage{amscd}
\usepackage{epsfig}
\usepackage{amsmath}
\usepackage{amssymb}

\newtheorem{definition}{Definition}

\newtheorem{conjecture}{Conjecture}

\begin{document}
\bibliographystyle{plain}

\title{Persistent Chaos in High Dimensions}
\author{D. J. Albers}
\email{albers@cse.ucdavis.edu}
\affiliation{Max Plank Institute for Mathematics in the Sciences,
  Leipzig 04103, Germany}
\affiliation{Computational Science and Engineering Center and Physics Department,
University of California, Davis, One Shields Ave, Davis CA 95616}
\affiliation{Physics Department, University of Wisconsin, Madison, WI 53706}
\affiliation{Santa Fe Institute, 1399 Hyde Park Road, Santa Fe, NM 87501}

\author{J. C. Sprott}
\email{sprott@physics.wisc.edu}
\affiliation{Physics Department, University of Wisconsin, Madison, WI 53706}

\author{J. P. Crutchfield}
\email{chaos@cse.ucdavis.edu}
\affiliation{Computational Science and Engineering Center and Physics Department,
University of California, Davis, One Shields Ave, Davis CA 95616}
\affiliation{Santa Fe Institute, 1399 Hyde Park Road, Santa Fe, NM 87501}

\date{\today}

\begin{abstract}
An extensive statistical survey of universal approximators shows that
as the dimension of a typical dissipative dynamical system is
increased, the number of positive Lyapunov exponents increases monotonically
and the number of parameter windows with periodic behavior decreases. A subset
of parameter space remains in which topological change induced by small
parameter variation is very common. It turns out, however, that if the system's
dimension is sufficiently high, this inevitable, and expected, topological
change is never catastrophic, in the sense chaotic behavior is preserved.
One concludes that deterministic chaos is persistent in high dimensions.
\end{abstract}

\preprint{Santa Fe Institute Working Paper 05-04-011}
\preprint{arxiv.org/abs/nlin/0504040}

\maketitle


Physical theory attempts to describe and predict the natural world by
expressing observed behavior and the governing balance of forces formally
in mathematical models---models that can only be approximate representations.
Empirically, many natural phenomena persist even when control parameters
and external conditions vary. For example, the essential character of fully
developed fluid turbulence is little affected if one slightly changes the
energy flux that drives it or if a small dent is made in the containing
vessel's wall. In building a theory of a system exhibiting this kind of
dynamical persistence, one hopes that, despite its approximations,
one's model also has this persistence.

\vspace{-0.025in}

A century of analyzing nonlinear dynamical systems, however, has lead to an
apparent inconsistency with this goal. Since the days of Poincar\'e's
development of qualitative dynamics, mathematicians and physicists have
probed differential equations to test their solutions for different kinds of
stability. Poincar\'e's discovery of deterministic chaos \cite{poincare_thesis}
demonstrated that at the most detailed level, there was inherent instability of
system solutions: change the initial condition only slightly and one finds
a different state-space trajectory develops rapidly. Later studies showed
that there was also an instability in behavior if the equations or parameters
were changed only slightly \cite{newhousewild1,Farm85a}. Even arbitrarily
small functional perturbations to the governing dynamic leads to radical
changes in behavior---from unpredictable to predictable behavior, for example.
The overall conclusion has been that nonlinear, chaotic systems are exquisitely
sensitive, amplifying arbitrarily small variations in initial and boundary
conditions and parameters to macroscopic scales.


How can one reconcile this with the observed fact of dynamical persistence
in many large-scale systems? We take \emph{dynamical persistence} to mean
that a behavior type ---e.g., equilibrium, oscillation, chaos---does
\emph{not} change with functional perturbation or parameter variation.
Here we describe the results of a Monte Carlo survey which empirically
demonstrate that chaos is dynamically persistent if the dimension of a
nonlinear system is sufficiently high. More constructively, we argue that
a particular mechanism is responsible for persistent chaos.

\vspace{-0.03in}

Specifically, the survey shows that in large-scale systems dynamical
sensitivity---when defined as breaking topological equivalences associated
with structural stability \cite{maness}, ergodicity \cite{pughshubAMS}, and
statistical stability \cite{alves_viana_stat_stab}---is typically benign and
does not affect behavior types. Naturally, drastic changes in the invariant
measure yield different observed dynamics, but the results indicate that
this becomes increasingly less probable. Moreover, the instability
associated with deterministic chaos dominates high-dimensional dynamical
systems, except at extreme parameter settings.

Much of the
intuition and motivation for our investigation comes from the analytical
results found in abstract dynamical systems theory, but our construction and
conclusions highlight a distinct difference. Said most simply, the number of
dimensions of the dynamical system matters. That is, there is a qualitative
difference between common behaviors in high- and low-dimensional dynamical
systems. Beyond giving empirical evidence to support these conclusions, we
introduce a definition of persistent chaos that suggests an alternative
approach to the long-standing questions of dynamical stability and offer a
mathematical conjecture on the mechanism underlying persistent chaos in
high dimensions.

\vspace{-0.025in}

The spectrum of Lyapunov characteristic exponents (LCE) \cite{benn2} will be
our primary tool for analyzing and identifying behavior types since there
is an equivalence between the number of negative and positive Lyapunov
exponents and the number of global stable and unstable manifolds,
respectively---structures that organize the state space and constrain
behavior \cite{ruellehilbert}. Therefore, in referring to topological
variation here we mean a change in the number of positive Lyapunov exponents.

\vspace{-0.02in}

In order to give a complete representation of the space of all systems, we
investigate typical behaviors in high dimensions using a class of dynamical
systems that are known to be \emph{universal function approximators}.
(They are universal in that they approximate arbitrarily closely any $C^r$
mapping, cf. \cite{hor2}.) These are single-layer neural networks of the form
\vspace{-0.1in}
\begin{equation}
x_{t} = \beta_0
      + \sum_{i=1}^{n}{{\beta}_i \tanh s \left( {\omega}_{i0}
          + \sum_{j=1}^{d}{{\omega}_{ij} x_{t-j} } \right)} ~,
\label{equation:net1}
\end{equation}
\vspace{-0.03in}
which are maps from $R^{d}$ to $R$. Here $n$ is the number of hidden units
(neurons), $d$ the number of time lags which determines the system's input
(embedding) dimension, and $s$ a scaling factor for the connection weights
$w_{ij}$. The initial condition is $({x}_{1}, x_{2}, \ldots, x_{d})$ and the
state at time $t$ is $({x}_{t}, x_{t+1}, \ldots, x_{t+d-1})$. The approximation
theorems of Ref. \cite{hor2} and time-series embedding results
of Ref. \cite{embedology} establish an equivalence between these neural
networks and general dynamical systems; cf. \cite{hypviolation}.

\vspace{-0.02in}

In the Monte Carlo survey we sample the $(n(d+1)+1)$-dimensional parameter
space taking (i) $\beta_{i} \in [0,1]$ uniformly distributed and rescaled
to satisfy $\sum_{i=1}^{n}{{{\beta}_{i}^{2}}} = n$, (ii) ${w}_{ij}$ as
normally distributed with zero mean and unit variance (which is adjusted
with $s$), and (iii) the initial $x_j \in [-1,1]$ as uniform. These
distributions---denoted $m_{\beta}$, $m_{w}(s)$, and $m_I$---form a product
measure on the space of parameters and initial conditions---the survey's
results then are statistical estimates with respect to this product measure.
We use the $s$ parameter as the primary control as it gives the magnitude
of the argument of $\tanh(x)$.  When $x \approx 0$ Eq. (\ref{equation:net1})
is linear and one finds fixed points and limit cycles; when $|x| \gg 1$
the output is binary and one finds $2^n$ different periodic states;
and for $|x|$ between these extremes we find the nonlinear behavior
we will focus on.





\vspace{-0.025in}

We define \emph{persistent chaos} in terms of the LCE spectrum as follows:
\vspace{-0.09in}
\begin{definition}[Degree-$p$ Persistent Chaos]
\label{definition:robustchaos}
Assume a discrete-time map $f$ that takes a compact set to itself. The map
has persistent chaos of degree-$p$ if there exists an open subset $U$
of parameter space, such that, for all $\xi \in U$ and a given open set
$\mathcal{O}$ of initial conditions, $f|_\xi$ retains $p \geq 1$ positive
Lyapunov exponents.
\end{definition}
\vspace{-0.09in}
Persistent chaos (\emph{$p$-chaos}) of degree $p$ is our notion of dynamical
equivalence on an open set of parameter space. This differs from that of a
robust chaotic attractor of Refs. \cite{unimodalrobust} and
\cite{yorkerobustchaos}, for example, in that we do not require the attractor
to be unique on the subset $U$. This is an important distinction since,
physically, there is little evidence indicating that such strict forms of
uniqueness are present in many complex physical systems \cite{milnorkaneko}
and, technically, uniqueness is significantly more difficult to demonstrate.
Indeed, alternative dynamical equivalence frameworks---such as nonuniform
partial hyperbolicity \cite{pesinlebook}---were invented to circumvent
problems with nonuniqueness.

\vspace{-0.025in}

\begin{figure}[tbp]
\begin{center}
\epsfig{file=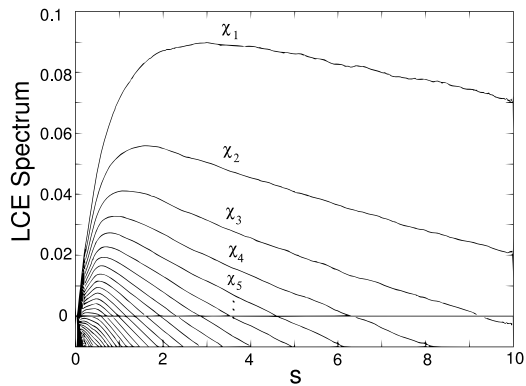, height=5.0cm}
\end{center}
\caption{LCE spectrum as a function of scale factor $s$ for a network of
  $32$ neurons and $64$ dimensions.  ($15000$ total time-steps; $5000$
  initial time-steps to arrive on the attractor.)}
\label{fig:biglced64}
\end{figure}

\vspace{-0.02in}

Figure \ref{fig:biglced64} represents the typical scenario for the LCE
spectra\footnote{For a $d$-dimensional system the spectrum consists of $d$
LCEs: $\chi_1 \geq \chi_2 \geq \ldots \geq \chi_d$, where indexing gives
a monotonic ordering.} of the high-dimensional neural networks as a function
of $s$. Here \emph{typical} refers to what was observed in $> 99 \%$ of
the $15800$ networks with $n\geq 32$ and $d>32$ given $m_{\beta}$, $m_{w}(s)$, and
$m_I$ as defined earlier.  Important features to notice include the lack of
periodic windows and that the LCEs vary continuously with $s$ and have a single
maximum (up to numerical fluctuations). The survey also reveals that
the maximum number of positive Lyapunov exponents is approximately $d/4$
and the attractors' Kaplan-Yorke dimension is roughly $d/2$ \cite{mythesis}.
As $d$ increases, the length of the $s$-intervals between LCE zero-crossings
decreases as $\sim d^{-1.92}$. These properties contrast sharply with the
familiar low-dimensional scenarios where one typically
encounters a preponderance of stable behavior and periodic windows and the
LCEs vary in a discontinuous manner with control parameters. (A more complete
analysis of these observations is found in Ref. \cite{hypviolation}.)
Moreover, this scenario is different from low-entropy, spatially-extended
systems discussed in \cite{grassberger_predictibility} or
\cite{kaneko_map_lattice_book}.  In our systems, the effective
dimension is irreducibly high and the uniformity of the dynamics is more robust.

\vspace{-0.025in}

These observations complement those from a previous study of chaos in
neural-network continuous-time differential equations \cite{Sompolinsky_nn}.
There, a mean-field analysis, which assumes that inputs are statistically
independent (and which does not apply in the present case), also suggested
that chaos should be common in high dimensions. 

\vspace{-0.025in}

In light of Fig. \ref{fig:biglced64} and the fact that LCE zero crossings
become asymptotically dense ($|U_i|\sim d^{-1.92}$ for $i<20, d\leqq 128$)
we propose the following dynamical mechanism for persistent chaos. For a
finite but arbitrarily large number of dimensions along an $s$
interval---e.g., $s \in (2, 8)$---there is an asymptotically dense, always
countable sequence of parameter values that have an LCE transversally
crossing through zero.  Thus, a continuous path along an $s$-interval
yields inevitable, but noncatastrophic (i.e. $p > 1$) topological change.
This implies that when varying parameters, periodic and quasiperiodic
windows will not exist in chaotic regions of parameter space of dynamical
systems with a sufficiently large number of positive exponents (i.e., when
entropy rate is large). The lack of dense periodic and quasiperiodic windows
is a necessary condition for $p$-chaos.



\vspace{-0.025in}

To test this picture, we analyzed the existence of periodic and quasiperiodic
windows along $s \in (1, 4)$ in networks with $n=32$ and $d$ ranging from
$8$ to $128$ and with an ensemble of $700$ networks per $n$ and $d$. We
observed that (i) the mean fraction of networks with periodic and
quasiperiodic windows decreases like $\sim d^{-1.3}$, (ii) the mean number
of windows decreases like $\sim d^{-2}$, and (iii) the window lengths
increase linearly with increasing $d$. These observations are insensitive
to increments in $s$ as long as $\Delta s \leq 0.005$. As the dimension
increased above $64$ the only networks with periodic windows had windows
that persisted for most of the $s$-interval under consideration. That is,
as dimension was increased, periodic windows became increasingly rare.
When they were observed, however, they were neither small nor intermittent,
but instead dominated the dynamics.

\vspace{-0.025in}

To explore the full parameter space systematically, one can fix $s$ and
vary the weights with random perturbations of a given size.  We surveyed
networks with parameters varied in a $(n(d+1)+1)$-ball with its center
fixed at $s$ noting that the results are insensitive to $s$ variation in
the chaotic portion of the $s$-interval; i.e., nearby $s$ values yield
identical results. Similarly, the results are insensitive to perturbation
size---weight variation on scales ranging from $10^{-10}$ to $1$ yield
similar results.

\vspace{-0.025in}

\begin{figure}[tbp]
\begin{center}
\epsfig{file=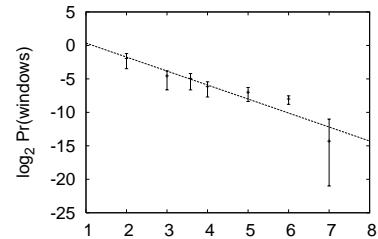, height=4.7cm, angle=270}
\caption{Log probability of periodic behavior versus log dimension for
  $700$ cases per $d$. Each case has all the weights perturbed on the
  order of $10^{-3}$; $100$ times per case.  The best-fit line is
  $\sim 1 /d^2$.
  }
\label{fig:probofwindows-high-d}
\end{center}
\end{figure}

The results are shown in Fig. \ref{fig:probofwindows-high-d}: the probability
of observing periodic windows decreases as the dimension increases. Each
data point corresponds to the probability of finding a system with a periodic
orbit among a set of $700$ networks at a given $n$ and $d$ and each perturbed
$100$ times. The range of weight perturbations was $10^{-3}$ with $s=3$. We
found that the probability of periodic networks decreases as $d^{-2}$. Thus,
as dimension increases the systems are far less likely to display periodic
windows and, as a consequence, become more persistently chaotic. 

\vspace{-0.025in}

While this is strong evidence for the disappearance of periodic windows
in parameter space with increasing dimension, a stronger argument follows
from our observation that the fraction of networks with windows decreases
less quickly ($\sim d^{-1.3}$) than the overall probability of windows
($\sim d^{-2}$). Thus,
periodic windows which do exist are concentrated in an ever-decreasing
fraction of networks and those with one periodic window
are more likely to have many periodic windows. 

\vspace{-0.025in}

To quantify the degree of $p$-chaos for an ensemble of mappings with our
construction consider the distribution of LCE's at a fixed $s$, noting
that $p$ depends on $s$.  If one takes $p$ to be the mean number of
positive LCE's, then $p$ increases monotonically like $\sim d/4$
\cite{mythesis} when $s$ is set to give the maximum number of positive
LCEs. If one considers $p$ to be the mean number of positive LCE's
minus $3$ standard deviations, a more conservative estimate, $p>0$
at $d=32$ and increases like $\sim 4 \log d$ at $s=3$ \cite{mythesis}.
We will refrain from arguing for a best definition of $p$ and simply
note that $p$ increases with $d$ monotonically. Together
with the periodic window data, this implies that while the parameter change
required to alter the absolute number of positive exponents decreases,
the perturbation required for \emph{all} the positive exponents to
vanish grows substantially.  Thus, the chance of falling into a periodic
window vanishes in proportion, and chaos becomes persistent over a
considerable portion of parameter space.   

\vspace{-0.025in}

\begin{figure}[tbp]
\epsfig{file=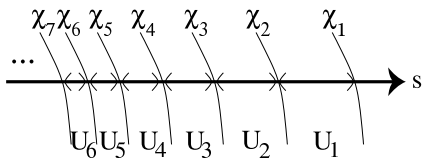, height=2.0cm}
\caption{Lyapunov spectrum versus network nonlinearity $s$: $U_i$'s are the
  open sets in parameter space where structural stability is believed to
  persist. The $|U_i|$ parameter intervals shrink like $\sim d^{-1.92}$ as
  the dimension increases.
  }
\label{fig:U_i}
\end{figure}

These observations and detailed analysis of $400$ four-dimensional
dynamical systems and $200$ $64$-dimensional dynamical systems, as
well as many of intermediate dimension, leads to the following view of
dynamic (topological) variation with parameter change. All of the
LCEs that become positive are negative for very small and very large
values of $s$---the LCE spectra exhibit a single maximum. As the
dimension $d$ is
increased, their variations decrease and their $s$-dependence becomes
smoother; recall Fig. \ref{fig:biglced64}. Moreover,
with increasing dimension the number of positive exponents increases
monotonically \cite{mythesis}. Finally, the distance between LCE
zero-crossings, above the maximum, decreases with dimension
as shown schematically in Fig. \ref{fig:U_i}.


\vspace{-0.025in}

Figure \ref{fig:U_i} is a graphical depiction of the hypothesized
persistence mechanism---a plot of
the $s$ axis transversally intersected by LCEs. In sufficiently high
dimensions, the subsets $U_i$ shrink and eventually
fall below any resolvability. The result, then, is twofold: one observes
continuous topological change (bifurcations), but this is never catastrophic.
One sees persistent chaos of varying degrees. The onset of sufficiently
high dimension for this to occur for our dynamical systems was observed
to be $d \geq 30$. These investigations lead to the following conjecture
for persistent chaos in high dimensions:
\vspace{-0.1in}
\begin{conjecture}
\label{conjecture:klceconjecture}
Assume $f$ as in Eq. (\ref{equation:net1}) and a sufficiently large number $d$
of dimensions and number $k=n(d+2)+1$ of parameters. There exists a large
\footnote{``Large'' depends on $d$: for $d=64$ we estimate a Lebesgue measure
$m(I_s)<20$; higher $d$ remains an open question.} Lebesgue
measurable set of $s \in R^1$ with respect to $m_{\beta}$, $m_{w}(s)$, and
$m_I$, for which chaos will be degree-$p$ persistent. Moreover,
$p \rightarrow \infty$ as $d \rightarrow \infty$.  
\end{conjecture}
\vspace{-0.1in}
Since our networks are universal function approximators, this behavior
should be observed in typical nonlinear high-dimensional dynamical systems. 

\vspace{-0.025in}

Two comments are in order. First, the existence of chaos as a persistent
behavior type depends on dimension. The subset of parameter space in which
chaos becomes persistent increases in size (with respect to Lebesgue
measure) as the dimension of the dynamical system increases. This is due
both to the increase in the number of positive LCEs (given a sufficient
increase in $n$) and to a decrease in the appearance of periodic windows.
Second, persistence is related to the number of (linearly independent)
parameters in the dynamical system. The number of neurons in the network
effectively controls the entropy rate \cite{manffra_thesis}---that is,
increasing the number of neurons increases the entropy rate, number of
positive exponents, and the maximum of the largest exponent. Increasing
$n$ simply increases the degree ($p$) of the persistent chaos, but the mechanism
for persistent chaos remains, due to the decreasing probability of periodic
windows. Networks with few parameters exhibit considerably less persistent
chaos.

\vspace{-0.025in}

In this way high entropy-rate systems are more persistent with respect to
functional \emph{and} parameter perturbations. This is in accord with a wide
range of experimental observations of such systems. Indeed, dynamical
persistence is not a novel experience; often hydrodynamic engineers and plasma
experimentalists expend much effort in attempts to eliminate persistent
chaos. Here we described a mechanism in which the dynamical persistence of
high-dimensional systems is retained under parameter perturbation,
despite the fact that stricter notions of dynamical equivalence are violated.
This sets the stage for more specific investigations of the statistical
topology of stable and unstable manifolds in high-dimensional
systems---investigations that, one hopes, will lead to predictive scaling
theories for observed macroscopic properties that are grounded in
microscopic dynamics.

\vspace{-0.03in}

We thank J. R. Albers, R. A. Bayliss, K. Burns, W. D. Dechert, D. Feldman,
J. Robbin, C. R. Shalizi, and J. Supanich for helpful discussions.
This work was partially supported at the Santa Fe Institute under
the Networks Dynamics Program funded by the Intel Corporation and under the
Computation, Dynamics, and Inference Program via SFI's core grants from the
National Science and MacArthur Foundations. Direct support for DJA was
provided by NSF grants DMR-9820816 and PHY-9910217 and DARPA
Agreement F30602-00-2-0583. 

\vspace{-0.3in}

\bibliography{dstexts,partialhyperbolicity,lyapunovexponents,neuralnetworks,nilpotency,topology,analysis,structuralstability,computation,me,physics,bifurcationtheory,probability,lorenz,srb,unimodal}

\end{document}